\documentclass[aps,prb,groupedaddress,amsmath,amssymb,superscriptaddress,showpacs,preprint]{revtex4-1}

\usepackage{graphicx} 
\usepackage{dcolumn} 
\usepackage{bm} 
\newcommand{\bmH}{\textbf{\textit{H}}}
\newcommand{\bmh}{\textbf{\textit{h}}}
\newcommand{\bmM}{\textbf{\textit{M}}}
\newcommand{\bmL}{\textbf{\textit{L}}}

\newcommand{\bmm}{\textbf{\textit{m}}}
\newcommand{\bml}{\textbf{\textit{l}}}
\newcommand{\bmd}{\textbf{\textit{d}}}

\usepackage{color}

\begin{document}

\title{Spectral dependence of photoinduced spin precession in DyFeO$_3$}

 \author{Ryugo Iida}
 \affiliation{Institute of Industrial Science, The University of
Tokyo, Tokyo 153-8505, Japan}
 \author{Takuya Satoh}
 \affiliation{Institute of Industrial Science, The University of
Tokyo, Tokyo 153-8505, Japan}
 \affiliation{PRESTO, Japan Science and Technology Agency, Saitama 332-0012, Japan}
 \author{Tsutomu Shimura}
 \affiliation{Institute of Industrial Science, The University of
Tokyo, Tokyo 153-8505, Japan}

 \author{Kazuo Kuroda}
 \affiliation{Institute of Industrial Science, The University of
Tokyo, Tokyo 153-8505, Japan}
 \author{B. A. Ivanov}
 \affiliation{Institute of Industrial Science, The University of
Tokyo, Tokyo 153-8505, Japan}
 \affiliation{Institute of Magnetism, Ukrainian Academy of Science,
Vernadskii Ave. 36B, 03142 Kiev, Ukraine}
 \author{Yusuke Tokunaga}
 \affiliation{Multiferroic
Project, ERATO, Japan Science and Technology Agency, Wako, Saitama
351-0198, Japan}
 \author{Yoshinori Tokura}
  \affiliation{Multiferroic Project, ERATO, Japan Science and
Technology Agency, Wako, Saitama 351-0198, Japan}
 \affiliation{Cross-Correlated Materials Research Group, ASI, RIKEN,
Saitama 351-0198, Japan}
 \affiliation{Department of Applied Physics,
The University of Tokyo, Tokyo 113-8656, Japan}

\date{\today}

\begin{abstract}
Spin precession was nonthermally induced by an ultrashort laser pulse in
orthoferrite DyFeO$_3$ with a pump--probe technique. Both circularly
and linearly polarized pulses led to spin precessions; these
phenomena are interpreted as the inverse Faraday effect and the
inverse Cotton--Mouton effect, respectively. For both cases, the
same mode of spin precession was excited; the precession frequencies
and polarization were the same, but the phases of oscillations were
different. We have shown theoretically and experimentally that the
analysis of phases can distinguish between these two mechanisms. We
have demonstrated experimentally that in the visible region, the
inverse Faraday effect was dominant, whereas the inverse
Cotton--Mouton effect became relatively prominent in the
near-infrared region.
\end{abstract}


\date{
\today}
\maketitle


\section{INTRODUCTION}

Magnetization switching triggered by femtosecond laser pulses has been studied in recent years.
Ultrafast demagnetization in ferromagnetic metals and semiconductors has also been reported.\cite{Beaurepaire,Wang}
These phenomena show thermal magnetic switching with light pulses on picosecond time scales.\cite{Koopmans}
However, heat-assisted spin reorientation is relatively slow because of the thermal diffusion time.

A light pulse with a certain polarization nonthermally modifies the
electron spin state.\cite{Ju,Dalla} Recently, it has been reported
that spin precession is induced by a circularly polarized pulse in
antiferromagnetic (AFM) DyFeO$_3$ with weak ferromagnetic (FM)
moment.\cite{Kimel1} The phase of spin precession changes by 180$^\circ$ on reversal of the pump helicity. The interpretation of this
phenomenon is that an effective magnetic field pulse parallel to the pump
wave vector is induced by the circularly polarized light pulse,
giving rise to the precession. The magnetic field generation effect
is referred to as the inverse Faraday effect (IFE). The same effect
has also been observed even in pure AFM NiO with no net magnetic
moment in the ground state.\cite{Satoh1} The resonance frequencies of
AFM materials reach the terahertz range, which is several orders of
magnitude higher than that of FM materials. For that reason, AFM materials
attract much attention in the context of ultrafast spin
control.\cite{Satoh1,Dodge,Kimel2,Zhao,Kimel3,Nishitani,Yamaguchi,Kampfrath,Higuchi} Spin precession is also observed
with a linearly polarized pump pulse, in particular, a pulse
polarized in a direction non-parallel to the crystal axes. This
phenomenon is called the inverse Cotton--Mouton effect
(ICME).\cite{Kalashnikova1,Kalashnikova2} A detailed review of these
phenomena can be found in Ref.~\onlinecite{Kirilyuk+RMPh}.

The ultrafast IFE and ICME are interpreted as impulsive stimulated Raman scattering (ISRS).\cite{Kimel4,Yan,Merlin}
An electron in the ground state is excited by the pump pulse into a virtual state, which changes the orbital momentum of the electron.
The nonzero orbital momentum flips the electron spin with spin-orbit coupling in the virtual state.
The excited electron radiates a photon and transits to the final state.
The energy gap between the final and ground states corresponds to the spin precession energy.

ISRS is a modulation of the dielectric permittivity by the pump pulse and should be dependent on the properties of the pulse, such as its polarization, wavelength, and fluence.
Therefore, examining the dependence of the photoinduced spin precession on these properties will help us to understand the ISRS mechanism.
In particular, it is not obvious how the pump photon energy influences spin precession.
An action spectrum of photoinduced spin precession should indicate the relation between the optical excited state and the spin precession via ISRS.

In the majority of previous publications, the excitation of
spin oscillations by ultrashort laser pulses was associated with IFE
and ICME separately. In the present work, we report spin precession
induced via ISRS as functions of the pump pulse polarization and
wavelength. We found that both effects, IFE and ICME are working in
the same way, exciting the same mode of spin precession. The phases
of the spin precession via IFE and ICME differ by 90$^\circ$,
allowing the two effects to be distinguished. We found an essential
dependence of the phase on the pump wavelength and demonstrated that
the IFE and ICME are dominating effects in different spectral
regions, in the visible region and in the near-infrared region,
respectively. Thus, the analysis of the phase difference of the spin
precession reveals the mechanism of ISRS.

\section{PHYSICAL PROPERTIES}

\subsection{Crystallographic and magnetic properties}

\begin{figure*}[tbp]
 \begin{center}
 \includegraphics[width=16cm,clip]{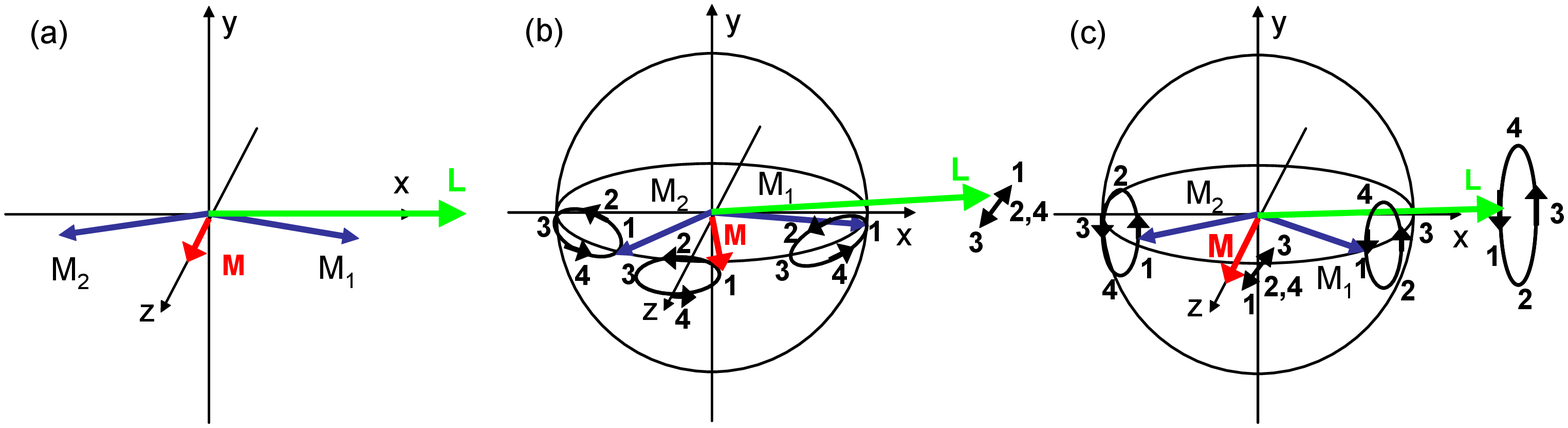}
 \end{center}
 \caption{(a) The static magnetic structure of DyFeO$_3$, with the four
   Fe$^{3+}$ spins regarded as satisfying $\bmM_1 \simeq \bmM_3$ and $\bmM_2 \simeq \bmM_4$. (b) Quasi-ferromagnetic and (c) quasi-antiferromagnetic spin precession.}
 \label{fig:fig1}
\end{figure*}

DyFeO$_3$ is a rare-earth orthoferrite and crystallizes in an
orthorhombic structure $D^{16}_{2h}$ $(Pbnm)$.\cite{Wijn} Spins
of the Dy$^{3+}$ ions are not ordered above 4~K. Four Fe$^{3+}$ ions
occupy positions (1/2,~0,~0), (1/2,~0,~1/2), (0,~1/2,~1/2), and
(0,~1/2,~0) in the unit cell. In the exchange
approximation, the arrangement of their magnetic
moments, $\bmM_1$, $\bmM_2$, $\bmM_3$, $\bmM_4$, corresponds to one of the
 four patterns $G_i:M_{1i}=-M_{2i}=M_{3i}=-M_{4i}$,
$F_i:M_{1i}=M_{2i}=M_{3i}=M_{4i}$,
$A_i:M_{1i}=-M_{2i}=-M_{3i}=M_{4i}$, and
$C_i:M_{1i}=M_{2i}=-M_{3i}=-M_{4i}$ $(i=x,$ $y,$ $z)$. DyFeO$_3$
crystal has the spin arrangement $\Gamma _4(G_x A_y F_z)$ and belongs to the
magnetic point group $m'm'm$ above the Morin point and below the
N\'{e}el temperature, at 37~K $<T<$ $T_N=645$~K.\cite{Gorodetsky,Tokunaga,Baryakhtar,Treves,White1} Because of the
superexchange interaction, the spins are almost completely arranged
antiferromagnetically along the $x$-axis. Due to the
Dzyaloshinskii--Moriya interaction, all spins tilt by about
$0.5^\circ$ toward the $z$-axis.\cite{Dzyaloshinskii,Moriya} Usually
the conditions $\bmM_1 \sim \bmM_3$ and $\bmM_2 \sim \bmM_4$ are
valid and a simpler model with just two different sublattice
magnetic moments, $\bmM_1$ and $\bmM_2$, with $|\bmM_1| = |\bmM_2| =
M_0$, can be employed.\cite{Baryakhtar,Herrmann} In what follows, this two-sublattice
model will be used.  We denote the FM vector by $\bmM = \bmM_1 + \bmM_2$ and
the AFM vector by $\bmL = \bmM_1 - \bmM_2$. These vectors are
subject to constraints
\begin{equation}\label{constrain}
(\bmM \cdot \bmL) = 0, \ \bmM^2+\bmL^2=4M_0^2 .
\end{equation}
The dynamics of $\bmM (t)$ and $\bmL (t)$ is described by  Landau--Lifshitz equations\cite{Baryakhtar,Balbashov,Turov1}
\begin{align}
\frac{d\bmM (t)}{dt}=-\gamma \{ [\bmM (t)\times \bmH ^{\rm {eff}}]+[\bmL (t)\times \bmh ^{\rm {eff}}]\}, \label{eq:dmdt} \\
\frac{d\bmL (t)}{dt}=-\gamma \{ [\bmM (t)\times \bmh ^{\rm
{eff}}]+[\bmL (t)\times \bmH ^{\rm {eff}}]\}, \label{eq:dldt}
\end{align}
where $\gamma = g\mu_\mathrm{B}/\hbar$ ($>0$) is the gyromagnetic
constant, $\mu_\mathrm{B}$ is the modulus of the Bohr magneton, $g$
is the gyromagnetic ratio, $g \approx 2$ for orthoferrites,
and $\bmH ^{\rm{eff}}$ and $\bmh ^{\rm{eff}}$ are the effective
magnetic fields. Using the magnetic energy of an orthoferrite, the
effective fields are denoted as $\bmH ^{\rm{eff}} =-\partial
\mathcal{H}/\partial \bmM$ and $\bmh ^{\rm{eff}} =-\partial
\mathcal{H}/\partial \bmL$, where the Hamiltonian is given
by\cite{Turov2,Baryakhtar}
\begin{equation}
\mathcal{H}=\frac{A}{2}\bmM ^2 + \frac{p_1}{2}M_x^2 + \frac{p_3}{2}M_z^2 + \frac{q_1}{2}L_x^2 + \frac{q_3}{2}L_z^2 -\bmd \cdot (\bmM \times \bmL). \label{eq:hamil}
\end{equation}
The last term describes the Dzyaloshinskii--Moriya interaction and
$\bmd$ is parallel to the $y$-axis. Equations (\ref{eq:dmdt}),
(\ref{eq:dldt}), linearized above the ground state  determined by the energy (\ref{eq:hamil}),
 yield eigenmodes of
oscillations of the vectors $\bmM(t)$ and $\bmL(t)$. These spin
precession modes for the $\Gamma _4$ ground state with the equilibrium values of $M_z \neq 0$ and
$L_x \neq 0$ (see Fig.~\ref{fig:fig1}(a))
are described as follows

with as following:
\begin{align}
\bmM(t) = M_z\hat{\textbf{z}} +\bmm(t), \\
\bmL(t) = L_x\hat{\textbf{x}} +\bml(t).
\end{align}
where $\hat{\textbf{z}}$ and $\hat{\textbf{x}}$ are unit vectors
parallel to the $z$-axis and $x$-axis, respectively, and
the variables
 $\bmm(t)$ and
$\bml(t)$ correspond to two eigenfrequency modes, as shown in
Fig.~\ref{fig:fig1}(b) and (c). The components
$m_x$, $m_y$ and $l_z$ oscillate at the
quasi-ferromagnetic resonance (F-mode) with the angular frequency
$\omega _{\rm F}$. On the other hand, $l_x$, $l_y$, and $m_z$
oscillate at the quasi-antiferromagnetic resonance (AF-mode) with the
angular frequency $\omega _{\rm AF}$.\cite{Herrmann} Those
 resonance frequencies are given by\cite{Turov2,Cinader,Koshizuka}
\begin{align}
\omega _{\rm F}=2\gamma M_0\sqrt{A(q_3-q_1)}, \\
\omega _{\rm AF}=2\gamma M_0\sqrt{d^2-Aq_1},
\end{align}
where $M_0=|\bmM_1|=|\bmM_2|$, and the anisotropy constants $p_1$ and $p_3$ are
omitted, because their contribution to frequencies is negligible.
(It is worth to note here, that the exchange-relativistic constant of the
Dzyaloshinskii--Moriya interaction has the same order of magnitude
as a square root of the product of exchange and relativistic constants like $\sqrt{A|q_1|}$
and thus should be kept in the above expressions.)
The orbits of the
spin precession and the temporal response of $\bmM$ and $\bmL$ are
different in the two modes.

\subsection{Optical and magneto-optical properties}

DyFeO$_3$ has the \textit{d-d} transitions~$^6A_{1g} \rightarrow $~$^4E_g,$~$^4A_{1g}$ centered at the wavelength of 500~nm, $^6A_{1g} \rightarrow $~$^4T_{2g}$ at 700~nm, and $^6A_{1g} \rightarrow $~$^4T_{1g}$ at 1000~nm.\cite{Wood,Kahn}
This crystal is optically biaxial, so the components of the dielectric permittivity tensor are $\varepsilon _{xx}^0 \not= \varepsilon _{yy}^0 \not=\varepsilon _{zz}^0$.
The birefringence $\rho $ stems from the difference in the refractive indices, such as $\Delta n_{xy}=n_x-n_y$~$(\rho =2\pi \Delta n/\lambda)$.
On the other hand, the magnetization $M_z$ leads to the Faraday rotation for the light propagating along the $z$-axis.
The Faraday effect is much smaller than the effect of birefringence.
In DyFeO$_3$, the Faraday rotation $\phi $ and the birefringence per unit length are $\phi = 1.6\times 10^3$ deg/cm and $\rho = 1.2 \times 10^5$ deg/cm, respectively, at 800~nm.\cite{Tabor1,Chetkin,Tabor2}

\subsection{Interaction of the light pulse and the medium}

The interaction of the magnetic medium and transmitting light is
described by the dielectric permittivity tensor
$\varepsilon _{ij}$.\cite{Smolenskii,LL_book} By virtue of the
Onsager principle, if absorption is negligible, $\varepsilon _{ij}$
can be divided into antisymmetric and symmetric parts, $(\varepsilon
_{ij}^a = -\varepsilon _{ji}^a)$ and $(\varepsilon _{ij}^s =
\varepsilon _{ji}^s)$, with real and imaginary components,
respectively. For a transparent medium, the tensor components can be written in
the following general form
\begin{equation}
\varepsilon _{ij} = \varepsilon _{ij}^{(0)}+i f_{ijk}M_k+
i g_{ijk}L_k+a_{ijkl}M_kM_l+b_{ijkl}L_kL_l+c_{ijkl}M_kL_l,
\end{equation}
where $\varepsilon _{ij}^{(0)}$ is a magnetization-independent term
having a symmetric part only. By taking into account the symmetry of orthoferrite, the terms in the first line
(except the $\varepsilon _{ij}^{(0)}$) represent the antisymmetric
part of  $\varepsilon _{ij}$, and the terms in the second line
describe the spin-dependent symmetric part of the permittivity
tensor. The symmetry of the fourth rank tensors $a_{ijkl}$,
$b_{ijkl}$, and $c_{ijkl}$ is determined by the magnetic and crystal
point groups, and $f_{ijk}$ and $g_{ijk}$ are the third rank
tensors, antisymmetric over the first pair of indices, e.g.,
$f_{ijk}=-f_{jik}$. Tensors  $c_{ijkl}$ and $g_{ijk}$  originate
from the Dzyaloshinskii--Moriya interaction. The Hamiltonian of the
interaction between the light pulse and the medium in SI unit is\cite{LL_book}
\begin{equation}
\mathcal{H}_{\rm int}=\frac{\varepsilon _{ij}}{4}\mathcal{E}_i(t)\mathcal{E}_j^*(t) \label{eq:hamiint2},
\end{equation}
where $\mathcal{E}_i(t)$ is the time-dependent amplitude of the
light in the pulse. A circularly polarized pulse propagating along
the $z$-axis can be described in the form $(\mathcal{E}_x(t),
\mathcal{E}_y(t))=\frac{\mathcal{E}(t)}{\sqrt{2}}(1, \pm i)$, where
the $\pm $ indicate the opposite senses of helicity. A linearly
polarized pulse with the polarization inclined on at an angle
$\theta $ with respect to the $x$-axis can be described in the form
$(\mathcal{E}_x(t), \mathcal{E}_y(t))=\mathcal{E}(t)(\cos\theta
,\sin\theta )$. Then a straightforward calculation gives the
Hamiltonian of the interaction  with the medium of the form
\begin{equation}
\mathcal{H}_{\rm int}^{\sigma ^{\pm }}= \frac{1}{8}\mathcal{E}(t)\mathcal{E}^*(t) \left( \varepsilon_{xx}^s+\varepsilon _{yy}^s\mp 2i\varepsilon_{xy}^a\right) \label{eq:hamiint3},
\end{equation}
\begin{equation}
\mathcal{H}_{\rm int}^{\rm lin}= \frac{1}{4}\mathcal{E}(t)
\mathcal{E}^*(t) \left(\varepsilon _{xx}^s\cos ^2\theta +\varepsilon
_{yy}^s\sin ^2\theta +\varepsilon _{xy}^s\sin 2\theta \right)
\label{eq:hamiint4},
\end{equation}
for circularly and linearly polarized
pulses,
respectively.
Nonzero components of the tensors $\varepsilon _{ij}^s$ and $\varepsilon _{ij}^a$ are listed in Table \ref{tab:tensor}.\cite{Eremenko,Zyezdin}

When a pulse is incident on a medium, the interaction between the pulse and the medium is given by Eqs. (\ref{eq:hamiint2})--(\ref{eq:hamiint4}).
\begin{table*}
 \caption{The dielectric permittivity tensor $\varepsilon _{ij}$ as a function of magnetic component. Modulation of the dielectric permittivity in F-mode and AF-mode is shown in columns 3 and 4, respectively. No external magnetic field is present. $\varepsilon _{ij}^a = -\varepsilon _{ji}^a$ and $\varepsilon _{ij}^s = \varepsilon _{ji}^s$}
 \label{tab:tensor}
 \begin{center}
 \begin{tabular}{|l|c|c|c|}
 \hline
 Tensor & Static & F-mode & AF-mode \\
 element & (\bmm (t)=0, \bml (t)=0) & ($m_x, m_y, l_z\not=0$) & ($l_x, l_y, m_z\not=0$) \\ \hline
 $\varepsilon _{xx}^s$ & $a_{xxzz}M_z^2+b_{xxxx}L_x^2$ & 0 & $(2a_{xxzz}M_z+c_{xxzx}L_x)m_z$ \\
  & $~~~~~~~~~+c_{xxzx}M_zL_x$ & & $~~~+(2b_{xxxx}L_x+c_{xxzx}M_z)l_x$ \\ \hline
 $\varepsilon _{yy}^s$ & $a_{yyzz}M_z^2+b_{yyxx}L_x^2$ & 0 & $(2a_{yyzz}M_z+c_{yyzx}L_x)m_z$ \\
  & $~~~~~~~~~+c_{yyzx}M_zL_x$ & & $~~~+(2b_{yyxx}L_x+c_{yyzx}M_z)l_x$ \\ \hline
 $\varepsilon _{zz}^s$ & $a_{zzzz}M_z^2+b_{zzxx}L_x^2$ & 0 & $(2a_{zzzz}M_z+c_{zzzx}L_x)m_z$ \\
  & $~~~~~~~~~+c_{zzzx}M_zL_x$ & & $~~~+(2b_{zzxx}L_x+c_{zzzx}M_z)l_x$ \\ \hline
 $\varepsilon _{xy}^s$ & 0 & 0 & $(2b_{xyxy}L_x+c_{xyzy}M_z)l_y$ \\ \hline
 $\varepsilon _{zx}^s$ & 0 & $(2a_{zxxz}M_z+c_{zxxx}L_x)m_x$ & 0 \\
  & & $~~~+(2b_{zxxz}L_x+c_{zxzz}M_z)l_z$ &  \\ \hline
 $\varepsilon _{yz}^s$ & 0 & $(2a_{yzyz}M_z+c_{yzyx}L_x)m_y$ & 0 \\ \hline
 $\varepsilon _{xy}^a$ & $if_{xyz}M_z+ig_{xyx}L_x$ & 0 & $if_{xyz}m_z+ig_{xyx}l_x$ \\ \hline
 $\varepsilon _{zx}^a$ & 0 & $if_{zxy}m_y$ & 0 \\ \hline
 $\varepsilon _{yz}^a$ & 0 & $if_{yzx}m_x+ig_{yzz}l_z$ & 0 \\
 \hline
 \end{tabular}
 \end{center}
\end{table*}
The incident pump pulse generates effective pulsed fields $\bmH
^{\rm{eff}} =-\partial \mathcal{H}_{\rm int}/\partial \bmM$ and
$\bmh ^{\rm{eff}} =-\partial \mathcal{H}_{\rm int}/\partial \bmL$.
Both effective fields are proportional to the intensity of the
light, $\mathcal{E}(t) \mathcal{E}^*(t)$. If the pulse duration
$\Delta$ is much shorter than the period of spin oscillations,
$\Delta \ll 1/\omega _{\rm F},1/\omega _{\rm AF}$, the real
pulse shape can be replaced by the Dirac delta function,
$\mathcal{E}(t) \mathcal{E}^*(t) \to I_0\delta(t)$, where $I_0=\int
\mathcal{E}(t) \mathcal{E}^*(t) dt$ is the integrated pulse
intensity.  The light-induced effective fields $\bmH
^{\rm{eff}}$ and $\bmh ^{\rm{eff}}$ can be regarded as being proportional to the delta
function $\delta (t)$ as well. For a light pulse propagating along the
$z$-axis, $\bmH ^{\rm{eff}}$ and $\bmh ^{\rm{eff}}$ generated by a
circularly polarized pulse are
\begin{equation}
\bmH^{{\rm eff,} \sigma ^{\pm}} =-\frac{I_0\delta (t)}{8}[2(a_{xxzz}+a_{yyzz})M_z+(c_{xxzx}+c_{yyzx})L_x\pm 2f_{xyz}]\hat{ \mathbf{z} }, \label{eq:h1cir}
\end{equation}
\begin{equation}
\bmh^{{\rm eff,} \sigma ^{\pm}} =-\frac{I_0\delta (t)}{8}[2(b_{xxxx}+b_{yyxx})L_x \\
+(c_{xxzx}+c_{yyzx})M_z\pm 2g_{xyx}]\hat{ \mathbf{x} }, \label{eq:h2cir}
\end{equation}
respectively. The phenomenon of generating these effective magnetic
fields is known as IFE.

For a magnetic field pulse of a short duration, the action of the
light-induced effective fields within the delta-function
approximation can be described as an instantaneous deviation of the
FM and AFM vectors, $\Delta \bmM = \bmM(t=+0)-\bmM(t=-0)$ and
$\Delta \bmL=\bmL(t=+0)-\bmL(t=-0)$, from their equilibrium
positions, $\bmM(t=-0)=M_z\hat{ \mathbf{z}}$ and
$\bmL(t=-0)=L_x\hat{ \mathbf{x}}$, respectively. After vanishing of
the pulsed effective field, the spins precess around the effective
fields corresponding to their equilibrium directions following the
Landau--Lifshitz equations, based on the Hamiltonian
\eqref{eq:hamil}. Thus the action of the pulse can be regarded as a
creation of some (non-equilibrium) initial conditions for the
Landau--Lifshitz equations. The deviation of the FM and AFM vectors
induced by the circularly polarized pulse is described by
\begin{equation}
\Delta \bmM^{\sigma ^{\pm}}=0, \label{eq:dmcir}
\end{equation}
\begin{multline}
\Delta \bmL^{\sigma ^{\pm}}=-\frac{\gamma I_0}{8}[2(b_{xxxx}+b_{yyxx}-a_{xxzz}-a_{yyzz})M_zL_x
+(c_{xxzx}+c_{yyzx})M_z^2 \\
-(c_{xxzx}+c_{yyzx})L_x^2
\mp 2(f_{xyz}L_x-g_{xyx}M_z)]\hat{ \mathbf{y} }. \label{eq:dlcir}
\end{multline}
Here, $\bmM$ is not affected by the effective field directly,
whereas $l_y$ of $\bmL(t)$ takes nonzero deviations. The resonance
mode with $l_y\not=0$ is AF-mode. In Fig.~\ref{fig:fig1}(c), two
spins $\bmM_1$ and $\bmM_2$, as well as their sum and difference
$\bmM$ and $\bmL$, move toward positions 2 or 4, and the spins
precess around their ground state directions. Because the spins
have only an $l_y$ variable component when the effective magnetic
field disappears, spin precession starts at position 2 or 4 (see
Fig.\ \ref{fig:fig1}(c)).

Similarly,  effective magnetic fields induced by a linearly polarized pulse are
\begin{equation}
\bmH^{\rm eff, lin} =-\frac{I_0\delta (t)}{4}[(2a_{xxzz}M_z+c_{xxzx}L_x)\cos ^2 \theta \\
+(2a_{yyzz}M_z+c_{yyzx}L_x)\sin ^2 \theta ]\hat{ \mathbf{z} }, \label{eq:h1lin}
\end{equation}
\begin{multline}
\bmh^{\rm eff, lin} =-\frac{I_0\delta (t)}{4}[ \{ (2b_{xxxx}L_x+c_{xxzx}M_z)\cos ^2 \theta
+(2b_{yyxx}L_x+c_{yyzx}M_z)\sin ^2 \theta \} \hat{ \mathbf{x} } \\
+(2b_{xyxy}L_x+c_{xyzy}M_z)\sin 2\theta \cdot \hat{ \mathbf{y} } ]. \label{eq:h2lin}
\end{multline}
These effective magnetic fields are induced via ICME.
The deviations of the FM and AFM vectors created by the effective field are
\begin{multline}
\Delta \bmM^{\rm lin}=\frac{\gamma I_0}{4}(2b_{xyxy}L_x^2+c_{xyzy}M_zL_x)\sin 2\theta \cdot \hat{ \mathbf{z} }, \label{eq:dmlin} \\
\end{multline}
\begin{multline}
\Delta \bmL^{\rm lin}=\frac{\gamma I_0}{4}[\{ (2b_{xxxx}M_zL_x+c_{xxzx}M_z^2
-2a_{xxzz}M_zL_x-c_{xxzx}L_x^2)\cos ^2 \theta \\
 +(2b_{yyxx}M_zL_x+c_{yyzx}M_z^2
-2a_{yyzz}M_zL_x-c_{yyzx}L_x^2)\sin ^2 \theta \} \hat{ \mathbf{y} } \\
-(2b_{xyxy}M_zL_x+c_{xyzy}M_z^2)\sin 2\theta \cdot \hat{ \mathbf{x}}]. \label{eq:dllin}
\end{multline}
Here, the components $m_z$ of $\bmM$ and $l_x$, $l_y$ of $\bmL$ are affected by the
effective field. This precession mode is also an AF-mode, but the initial
direction of the spin deviation differs from that for the circularly
polarized pulse case.

A pulse propagating along the $x$- or $y$- axis should trigger the spin precession with both F- and AF- modes.
The amplitude and the phase of the precession depends on the polarization of the pulse.

\section{EXPERIMENTAL RESULTS}

\subsection{Experimental setup}

\begin{figure}[htbp]
 \begin{center}
 \includegraphics[width=8.5cm,clip]{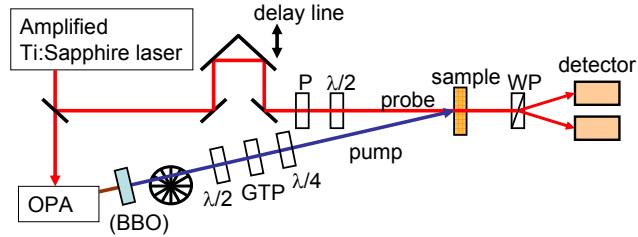}
 \end{center}
 \caption{The experimental setup geometry. BBO was used for frequency doubling of the pump pulse, if necessary. WP: Wollaston prism, GTP: Glan--Taylor prism, P: polarizer, $\lambda /2$: half-wave plate, $\lambda /4$: quarter-wave plate.}
 \label{fig:fig2}
\end{figure}

We studied photoinduced spin precession in DyFeO$_3$ using a pump--probe magneto-optical technique, as shown in Fig.~\ref{fig:fig2}.
DyFeO$_3$ single crystals were grown by the floating-zone method,
and the orientation of the faces were determined by back-reflection
x-ray Laue photographs.\cite{Tokunaga} Faces with a width of few
millimeters were mechanically polished. The sample thickness was
140~$\mu $m, except for a thickness dependent measurement. The
sample was placed in a cryostat at 77~K with no external magnetic
field. Optical pulses with a central wavelength of 790~nm, a
duration of 150~fs, and a repetition rate of 1~kHz were emitted from
an amplified Ti:sapphire laser. The beam was separated into two
beams by a beamsplitter. One was employed as the probe beam, and the
other was injected into an optical parametric amplifier (OPA), which
converted the incident beam to signal and idler beams, in the
wavelength ranges 1140--1580~nm and 1580--2570~nm, respectively.
Furthermore, the signal and idler beams were frequency-doubled with
a $\beta $-BaB$_2$O$_4$ (BBO) crystal if necessary. Then unwanted beams were cut by
color filters. The ranges of the pump wavelength were 600--750~nm
(second harmonic of the signal pulse), 850--1100~nm (second harmonic
of the idler pulse), and 1140--1500~nm (the signal pulse).

\begin{figure}[htbp]
 \begin{center}
\includegraphics[width=6cm,clip]{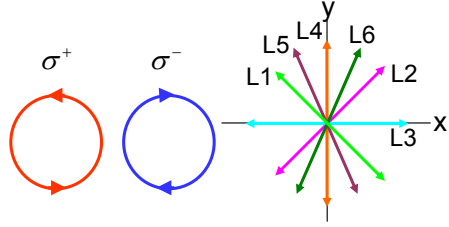}
 \end{center}
 \caption{Pulse polarizations. Circularly polarized pulses are denoted
 $\sigma ^{\pm }$. Linearly polarized pulses, denoted L1, L2, L3, L4, L5, and L6,
 were tilted at $-\pi /4, \pi /4, 0, \pi /2,$ and $\mp \alpha $ from the $x$-axis,
 respectively, where $\tan\alpha =2$.}
 \label{fig:fig3}
\end{figure}

Figure \ref{fig:fig3} illustrates the circular and linear
polarizations employed for the pump and probe pulses. 
Circularly polarized pulses are denoted as $\sigma ^{\pm }$. Linearly polarized
pulses, denoted L1, L2, L3, L4, L5, and L6, were tilted at $-\pi /4,
\pi /4, 0, \pi /2,$ and $\mp \alpha $ from the $x$-axis,
respectively, where $\tan\alpha =2$.

The fluence of the pump pulse was varied from 15 to 130 mJ/cm$^2$, depending on the wavelength.
The pump pulses were focused on the sample to spot sizes of 50--100 $\mu$m.
The probe pulses were linearly polarized and had a pulse fluence of 1 mJ/cm$^2$.
The probe beam was vertically incident on the surface of the sample, whereas the pump beam was incident at the angle of $7^\circ$.
The transmitted probe pulse was divided into two orthogonally polarized pulses by a Wollaston prism, and each pulse was detected with a Si photodiode.
The ratio of the signals from the detectors allowed us to determine the angle of the probe polarization.

\subsection{Dependence of the polarization rotation on the pump pulse polarization}

\begin{figure}[htbp]
 \begin{center}
 \includegraphics[width=8cm,clip]{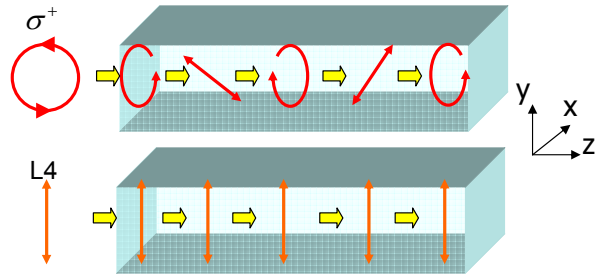}
 \end{center}
 \caption{Polarizations of the propagating pulses in the medium with birefringence.}
 \label{fig:fig4}
\end{figure}

Figure \ref{fig:fig4} illustrates the polarization of the
propagating pulses in the medium with birefringence. For the sake of
simplicity, we will discuss the picture of the light propagation
without taking the Faraday effect into account.
Pulses with circular polarization or linear polarization nonparallel
to the crystal axis are transformed, whereas pulses with linear
polarization parallel to the crystal axis, corresponding to the
normal modes of light in the media, are not.
Thus, for the pulses with general linear polarization or pulses
with circular polarization, the real and imaginary parts of
$E_xE_y$, which are responsible for the terms in Eqs.
(\ref{eq:h1cir})--(\ref{eq:dllin}) including $\sin2\theta $, and
$f_{xyz}$ and $g_{xyx}$, respectively, will oscillate in space along
the pulse propagation direction, while they remain uniform only for
pulses linearly polarized parallel to the crystalline axis.
Therefore, the effective magnetic field and spin precession
generated by IFE and ICME will be different at different positions
in the sample.

\begin{figure*}[htbp]
 \begin{center}
 \includegraphics[width=13cm,clip]{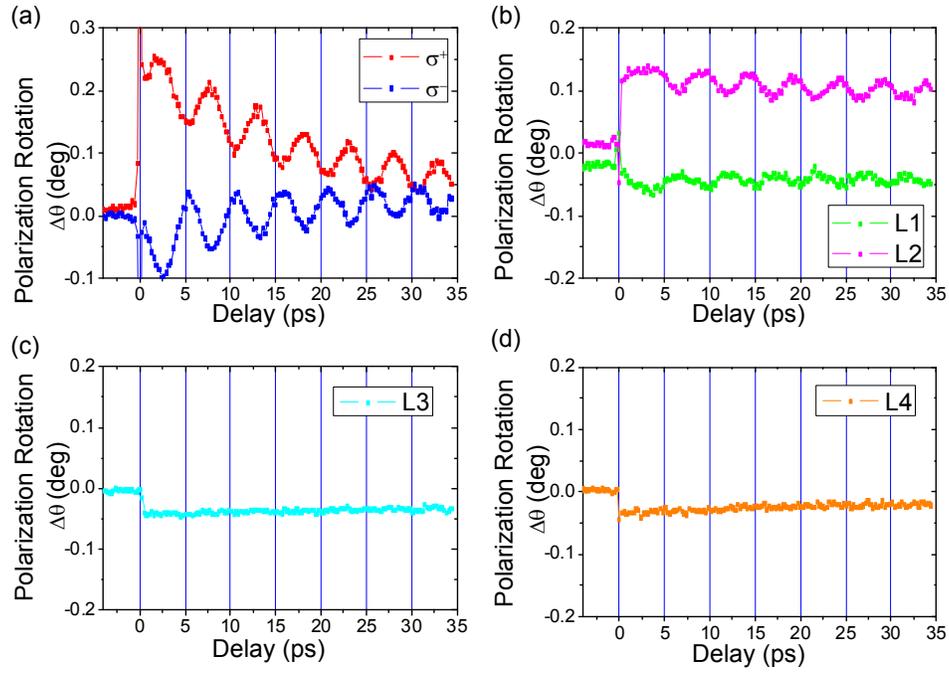}
 \end{center}
 \caption{Oscillation of the probe polarization $\theta (t)$ as a function of the time delay between the pump and probe pulses. Six types of probe polarizations were used: (a) circular polarization $\sigma ^{\pm }$, and linear polarizations tilted at (b) $\mp 45^\circ$, (c) $0^\circ$, and (d) $90^\circ$ with respect to the $x$-axis.}
 \label{fig:fig5}
\end{figure*}

Figure \ref{fig:fig5} shows the polarization rotation of the probe pulse as a
function of the delay time between the pump and probe pulses.
The pump wavelength was 1050 nm, and the polarizations were $\sigma ^{\pm }$, L1, L2, L3, and L4.
The probe polarization was L4.
When the pump polarizations were $\sigma ^{\pm }$, L1, and L2,
oscillation of the polarization rotation was observed. The frequency
of the oscillation was 210 GHz at the temperature $T=77$ K, in agreement with previous infrared and Raman
experiments.\cite{Balbashov,Koshizuka,White2} In Figs.
\ref{fig:fig5}(c) and (d), the pump pulses with polarizations L3 and
L4 did not induce oscillation of the probe polarization.
Polarizations $\sigma ^{\pm }$, L1, and L2 had $E_xE_y$ components,
but L3 and L4 did not, as shown in Fig. \ref{fig:fig4}.

\subsection{The influence of magnetization on the probe polarization}

Modulation of the dielectric permittivity leads to oscillation  of
the probe polarization in the sample. The origin of the probe
polarization change can be attributed to the Cotton--Mouton effect
and the Faraday effect. The Cotton--Mouton effect is magnetic linear
birefringence based on $\varepsilon _{xy}^s$, whereas the Faraday
effect is magnetic circular birefringence based on $\varepsilon_{xy}^a$.

\begin{figure}[htbp]
 \begin{center}
 \includegraphics[width=6cm,clip]{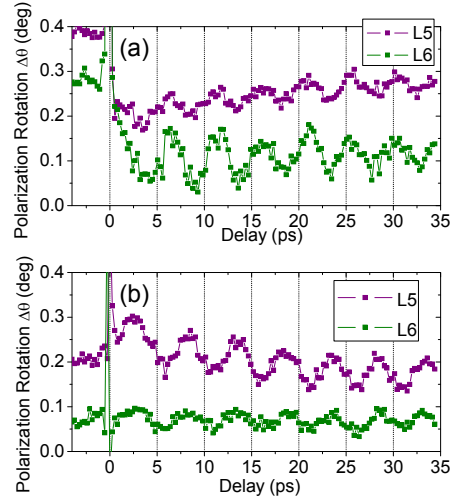}
 \end{center}
 \caption{Time-resolved pump-induced probe polarization $\theta (t)$. The pump
   polarization was circular, and the graph shows the shift of the probe
     polarization when changing the pump polarization from $\sigma ^+$ to $\sigma ^-$. The pump wavelengths were (a) 750~nm and (b) 1050~nm, and the probe polarizations were L5 and L6.}
 \label{fig:fig6}
\end{figure}

In order to identify the effect giving rise to the
polarization rotation as observed in Fig.~\ref{fig:fig5}, we set
$\sigma ^{\pm}$ for the pump polarization and L5 and L6 for the
probe polarization. For L5 and L6, the Faraday effect leads to
rotation of the probe polarization in the same direction for both
probe polarizations, whereas the Cotton--Mouton effect leads to
rotation in the opposite direction.
Therefore, the dominance of the rotation of the probe polarization can be distinguished.
Figure \ref{fig:fig6} shows that the polarization rotations of two probe pulses with polarizations L5 and L6 oscillated in the same direction.
This indicates that the contribution of the Faraday effect is dominant and that
of the Cotton--Mouton effect is negligible for the probe wavelength of 800~nm.
This is consistent with the fact that IFE is dominant for the pump wavelength of 800~nm (see below).

\subsection{Dependence of the polarization rotation on the pump wavelength}

The oscillation of the probe polarization originates from spin
precession.
Therefore, the phase of the oscillation indicates the direction of
an effective magnetic field induced by the pump beam. The dependence
of the effective magnetic field and reorientation of magnetization
on the pump wavelength gives information about the interaction of
the light pulse and the magnetic medium.

\begin{figure}[htbp]
 \begin{center}
 \includegraphics[width=8cm,clip]{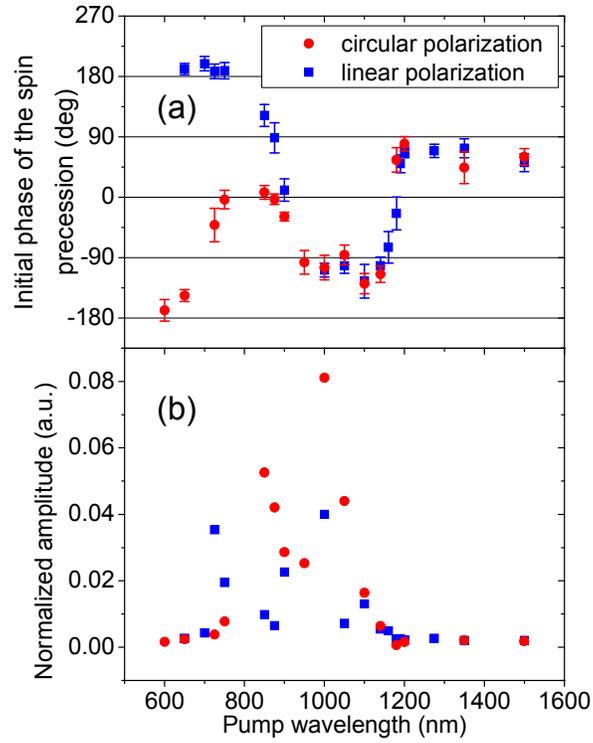}
 \end{center}
 \caption{(a) Initial phase and (b) amplitude of the oscillation of the
   polarization as a function of the pump wavelength. The amplitude is
   normalized by the pump fluence.}
 \label{fig:fig7}
\end{figure}

An experiment was performed with four types of pump polarizations, $\sigma ^\pm$, L1, and L2.
The differences between the oscillations for $\sigma ^+$ and $\sigma ^-$ and between those for L1 and L2 were measured.
Figure \ref{fig:fig7} (a) shows the initial phase $\xi $ of the oscillation of the probe polarization versus pump wavelength.
The oscillation is described by $\theta (t)=A\sin (\omega t+\xi )$ at $t>0$, where $A$ is the amplitude, $\omega $ is the angular frequency, and $\xi $ is the initial phase.
The initial phase was close to $0^\circ$ (or $\pm 180^\circ$), when
the pump wavelength was 800 nm. This is consistent with
Ref.~\onlinecite{Perroni}.

When the pump wavelength was between 1000~nm and 1100~nm, the
initial phase was closer to $\pm 90^\circ $. When the pump
wavelength was above 1200~nm, the initial phase was between
$0^\circ$ and $90^\circ$. By comparing two samples with thicknesses
of 140~$\mu $m and 170~$\mu $m, it was confirmed that the sample
thickness does not affect the phase shift (data not shown). Figure
\ref{fig:fig7} (b) represents the amplitude $A$ of the oscillation
as a function of the pump wavelength. The amplitude $A$ is
proportional to the pump fluence, thus justifying normalization of
the amplitude by the fluence. Because of the transformation of the
pulse polarization in the medium, the magnetic field differs at
different positions. Therefore, the amplitude $A$ was not simply
proportional to the magnitude of the generated magnetic field.
However, when the pump wavelength was from 700 nm ($^6A_1
\rightarrow $ $^4T_2$) to 1000 nm ($^6A_1 \rightarrow $ $^4T_1$),
the amplitude was larger than that of the other region in
Fig.~\ref{fig:fig7}~(a). This result suggests that the photoinduced
spin precession is related to the electron transition.

\subsection{Pump--probe measurement in (100) and (010) oriented crystals}

To determine all dielectric permittivities, we performed pump--probe
measurements in (100) and (010) oriented crystals. The pump
wavelength was 750~nm, and the crystal thickness in both cases was 100~$\mu$m. However, in contrast to the previous experiments,\cite{Kimel1}
oscillation of the polarization of neither F- nor AF-modes was observed in either propagation direction.

\subsection{The dependence of polarization rotation on temperature}

\begin{figure}[htbp]
 \begin{center}
  \includegraphics[width=6cm,clip]{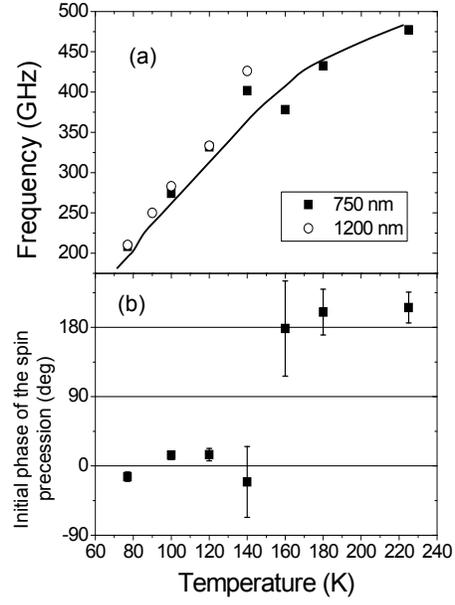}
 \end{center}
 \caption{(a) Temperature dependence of the induced spin precession
   frequency. Pump wavelengths were 750~nm and 1200~nm. The solid line
     shows the magnon frequency taken from Ref. \onlinecite{Balbashov}. (b)
   Temperature dependence of the initial spin precession phase for the pump wavelength of 750~nm.}
 \label{fig:fig8}
\end{figure}

It is well known that  magnon frequencies in orthoferrites
strongly depend on the temperature.\cite{Kimel1,Balbashov,Koshizuka,White2}
We measured the temperature dependence of
 the spin precession properties in DyFeO$_3$.
The frequencies of the oscillations for pump wavelengths of 750~nm
and 1200~nm are shown in Fig. \ref{fig:fig8} (a), in comparison with
previously reported spin precession.\cite{Balbashov} Our data show
excellent agreement with Refs.\ \onlinecite{Kimel1} and \onlinecite{Balbashov}, regardless of the pump wavelength. The frequency decreases with
approaching the Morin point $T_r=37$~K because of magnon softening
associated with the spin
reorientation.\cite{Balbashov,Koshizuka,White2} The temperature
dependence of the initial phase $\xi $ of the spin precession for pump wavelength of 750~nm is
shown in Fig. \ref{fig:fig8} (b). The initial phase was close to
0$^\circ$ or 180$^\circ$ with a jump at $T$=150~K.
It is worth to note  that at this temperature the frequencies of F-mode and
AF-mode become equal,
that is $A(q_3-q_1)=d^2-Aq_1$.
Furthermore, the energies of two domain walls with the spin rotation in (010) and (001) planes become equal at this point, which leads to the reconstruction of domain walls.\cite{Zalesskii}
However, we were not able to find the relation between the properties described above and the initial phase shift.

\section{DISCUSSION}

\subsection{Landau--Lifshitz equations}

According to the results of the previous section, the number of essential dielectric permittivity components can be reduced.
First, we found that pump--probe measurement in (100) and (010) oriented crystals revealed that $\varepsilon _{zz}^s$, $\varepsilon _{xz}^s$, $\varepsilon _{yz}^s$, $\varepsilon _{xz}^a$, and $\varepsilon _{yz}^a$ were negligible.
In addition, pump pulses with L3 and L4 polarizations did not trigger spin precession in the (001) oriented crystal in Fig. \ref{fig:fig5}.
Polarizations L3 and L4 had only electric field components $E_x$ and $E_y$, respectively.
Therefore, the terms containing $\cos ^2 \theta $ and $\sin ^2 \theta $ in Eqs. (\ref{eq:h1lin}), (\ref{eq:h2lin}) and (\ref{eq:dllin}) were also negligible.

Moreover, it has been reported that $f_{xyz}M_z$ and $g_{xyx}L_x$ are of
the same order of magnitude for orthoferrites.\cite{Woodfood1}
In contrast to that, for the AF-mode the ratio of $m_z$ and $l_x$ is
$|m_z/l_x|=|L_x/M_z|\simeq 100$. Thus, $f_{xyz}m_z \gg
g_{xyx}l_x$, and one can ignore the term $g_{xyx}l_x$. In addition,
Fig.~\ref{fig:fig6} indicates that the observed oscillation of the
polarization was dominated by the imaginary part of the dielectric
permittivity $\varepsilon
_{xy}^a=if_{xyz}(M_z+m_z)+ig_{xyx}(L_x+l_x)$. Because $f_{xyz}m_z
\gg g_{xyx}l_x$, the phase of $m_z$ corresponds mostly to that of
$\varepsilon _{xy}^a$ and that of the oscillation of the
polarization.

These findings simplify the dielectric permittivity tensor. In
Table \ref{tab:tensor}, the tensor elements in the AF-mode column
are proportional to $l_x$ and negligible, except for
$\varepsilon_{xy}^s=(b_{xyxy}L_x+c_{xyzy}M_z)l_y$ and
$\varepsilon_{xy}^a=if_{xyz}m_z$.
Here we suppose that a pulse is incident
on a (001) oriented crystal. We can simplify the effective magnetic
field and the dynamics of the magnetization induced by the circular
polarization:
\begin{equation}
\bmH^{{\rm eff,} \sigma ^{\pm}} =\mp \frac{I_0\delta (t) f_{xyz}}{4}\hat{ \mathbf{z} }, \label{eq:h1cir2}
\end{equation}
\begin{equation}
\bmh^{{\rm eff,} \sigma ^{\pm}} =0, \label{eq:h2cir2}
\end{equation}
\begin{equation}
\Delta \bmM^{\sigma ^{\pm}}=0, \label{eq:dmcir2}
\end{equation}
\begin{equation}
\Delta \bmL^{\sigma ^{\pm}}=\pm \frac{\gamma I_0 f_{xyz}L_x}{4}\hat{ \mathbf{y} }. \label{eq:dlcir2}
\end{equation}
In the case of the linear polarization  one in turn obtains:
\begin{equation}
\bmH^{\rm eff, lin} =0, \label{eq:h1lin2}
\end{equation}
\begin{equation}
\bmh^{\rm eff, lin} =-\frac{ I_0\delta
(t)}{4}(2b_{xyxy}L_x+c_{xyzy}M_z)\sin 2\theta \cdot \hat{ \mathbf{y}
}, \label{eq:h2lin2}
\end{equation}
\begin{equation}
\Delta \bmM^{\rm lin}=\frac{\gamma I_0}{4}(2b_{xyxy}L_x^2+c_{xyzy}M_zL_x)\sin 2\theta \cdot \hat{ \mathbf{z} }, \label{eq:dmlin2}
\end{equation}
\begin{equation}
\Delta \bmL^{\rm lin}=-\frac{\gamma I_0}{4}
(2b_{xyxy}M_zL_x+c_{xyzy}M_z^2)\sin 2\theta \cdot \hat{ \mathbf{x} }. \label{eq:dllin2}
\end{equation}
The second terms are much smaller than the first ones in Eqs. (\ref{eq:h2lin2}), (\ref{eq:dmlin2}) and (\ref{eq:dllin2}), respectively.
As a result, IFE and ICME are induced by the contributions of $\varepsilon ^a_{xy}$ and $\varepsilon ^s_{xy}$, respectively.

Equation (\ref{eq:dlcir2}) indicates that the circular polarization causes the AFM component $l_y$ and rotation torque of the AF-mode.
On the other hand, the FM component does not change in
Eq.~(\ref{eq:dmcir2}). As a result, IFE leads to oscillations proportional to
$\sin \omega _{\rm AF}t$.
On the other
hand, Eqs.\ (\ref{eq:dmlin2}) and (\ref{eq:dllin2}) indicate that
linear polarization causes components $m_z$ and $l_x$. As shown in
Fig.\ \ref{fig:fig1}, $m_z$ and
$l_x$ have the same phase,  so
ICME leads to oscillations proportional to $\cos \omega _{\rm AF}t$.
Therefore, the initial phases of $m_z$ excited by IFE and ICME
differ by 90$^\circ$.

Because the phase of $m_z$ is nearly equal to that of the
oscillation of the polarization, we can estimate the phase of spin
precession from the result in Fig. \ref{fig:fig7}. Since the
polarization of the pump pulse is transformed by birefringence, the
effective magnetic field and spin precession differ at different
positions in the medium, as shown in Fig. \ref{fig:fig4}. However,
if one of IFE and ICME is dominant and the other is negligible, the
time dependence of $m_z$ and the oscillation of the probe
polarization are proportional to $\sin \omega _{\rm AF}t$ or $\cos
\omega _{\rm AF}t$, respectively.

\subsection{Sigma model}

Nonlinear sigma model is a convenient tool for
the description of linear and
especially non-linear spin dynamics of antiferromagnets, see
Ref.~\onlinecite{Baryakhtar} for details.
It is based on the dynamical equation for the vector $\bm L$ only that is
of the second order in time
derivatives, whereas the vector $\bmM$
is a slave variable so it can be expressed through the vector $\bm L$
and its time derivative.
Recently, two alternative scenarios of laser-induced excitations of
spin oscillations in antiferromagnets have been discussed within the framework of this model.
The so-called \emph{inertial} mechanism
has been proposed for canted antiferromagnets and has been realized
experimentally for holmium orthoferrite.\cite{Kimel3} Within the
sigma-model approach, the inertial mechanism is associated with an
action of the laser-induced pulse of the magnetic field on the
vector $\bm L$ as a pulse of  force on the massive particle. In
this mechanism, the laser pulse creates  an initial value of the
time derivative, $d\bm L/dt$, that in principle can lead to quite
large deviations of the vector  $\bm L$ after the action of the pulse. In the
alternative mechanism, the \emph{time derivative} of the effective
magnetic field plays a role of the driving force, leading to an
initial deviation of the vector $\bm L$ from its equilibrium
direction.~\cite{Galkin,Satoh1} For this \emph{field-derivative
mechanism}, the amplitudes of spin deviations are expected to be
smaller than for inertial mechanism, but can be realized for any
antiferromagnet, even a purely compensated one. The latter mechanism has beed
observed experimentally in AFM nickel oxide, where the Dzyaloshinskii-Moriya
interaction is forbidden by symmetry.~\cite{Satoh1} It is
interesting to understand which mechanism is responsible for the
spin oscillations observed in the present work.

The Lagrangian density of the sigma model can be written as
follows:\cite{Baryakhtar}
\begin{equation}
\mathcal{L}=\frac{1 }{8\gamma^2 AM_0^2}\left( \frac{\partial
\bmL}{\partial t}\right) ^2 +
 \frac{1 }{4\gamma AM_0^2}\left( \bmH  \cdot \left( \frac{\partial \bmL}{\partial t} \times \bmL \right) \right)
 + \frac{1}{A} (\bmH \cdot (\bmL \times  \bmd)) - \mathcal{W}_a(\bmL )\,, \label{eq:lag}
\end{equation}
where $\bmH$ is the effective magnetic field and $\mathcal{W}_a(\bmL
)$ is the effective anisotropy  energy that includes  $L$-dependent
terms from the Hamiltonian \eqref{eq:hamil} and a
contribution from the Dzyaloshinskii-Moriya interaction, see Eq.~(\ref{eq:Weff})
below. The slave variable,  magnetic moment
$\bmM$, can be easily expressed via vector $\bmL$
and its time derivative,
\begin{equation}
\bmM = \frac{\bmL \times \bmd }{A} + \frac{\bmH L^2- \bmL \left(
\bmL \cdot \bmH \right) }{AL^2} +\frac{1}{\gamma A L^2}\left(
\frac{\partial \bmL}{\partial t} \times \bmL\right) \label{eq:M}\, ,
\end{equation}
where $L= |\bmL|$. Within the sigma-model approximation, the
length of the vector $\bmL$ should be treated as a constant,
$L_x^2+L_y^2+L_z^2=\mathrm{const}\simeq (2M_0)^2$. Thus, in the
linear approximation  $L_x \simeq 2M_0- (l_y^2+l_z^2)/4M_0$ and the
two components, $l_y$ and $l_z$ can be considered as independent
variables. It is in line with our experimental observation that the
component $l_x$ is completely negligible. The effective anisotropy
energy can be taken in the form
\begin{equation}
\mathcal{W}_a(\bmL )= \frac{1}{2}\left(q_3-q_1\right)l_z^2 +
\frac{1}{2}\left(\frac{d^2}{A}-q_1\right)l^2_y\,, \label{eq:Weff}
\end{equation}
where the additive constant is omitted. Free oscillations of the two components at
$\bmH=0$ correspond to two independent
magnon modes (F- and AF-modes), described by the following equations
\begin{eqnarray}
 \nonumber
  \frac{d^2l_z}{dt^2}&+& \omega^2_{\mathrm{F}}l_z =0,\,
  \bmm=\hat{\mathbf{x}}
  \frac{d}{A}l_z + \hat{\mathbf{y}} \frac{1}{2\gamma A M_0} \frac{dl_z}{dt}\,,  \\
  \frac{d^2l_y}{dt^2}&+& \omega^2_{\mathrm{AF}}l_y =0,\, \bmm=-\hat{\mathbf{z}}
  \frac{1}{2\gamma A M_0} \frac{dl_y}{dt}\,. \label{eq:Modes}
\end{eqnarray}

Now let us discuss the excitations of the modes by light pulses. The
interaction of the spin system with the light is described by the
Hamiltonian \eqref{eq:hamiint2}, that for the specific case of
circularly or linearly polarized light reads as
\eqref{eq:hamiint3} or \eqref{eq:hamiint4}, respectively. Within the
sigma-model approach, for different polarizations the interaction terms enter different
parts of the Lagrangian (\ref{eq:lag}): the circularly polarized
light contributes to the effective field
$\bmH=\bmH^{{\rm eff,} \sigma ^{\pm}}$, whereas the effect of the
linearly polarized light is described by the
time-dependent contribution
\begin{equation}
\delta \mathcal{W}_a(\bmL ,t)=\frac{1}{4}\mathcal{E}_i(t)
\mathcal{E}_j^*(t) b_{ijkl}L_kL_l,
\end{equation}
to the effective anisotropy energy $ \mathcal{W}_a(\bmL )$. Among
all these contributions to the Lagrangian, we need to find terms
linear on $l_y$ and $l_z$, which produce the ``driving force'', i.e., lead to a
non-zero right-hand side in the equations of motion~\eqref{eq:Modes}.

The light-induced effective field is directed along $z$-axis, and it
is easy to see that the term $(\bmH \cdot (\bmd \times \bmL ))$
gives no ``driving force'' contributions for both modes. The
gyroscopic term with $d\bml/dt$ provides such a term for $y$-component of the vector $\bml$, proportional to $H^{{\rm eff,}
\sigma ^{\pm}}_z L_x(dl_y/dt)$, but not for its $z$-component. Thus,
for the state of interest ($\bmL=L_x\hat{\mathbf{x}} $ in the
ground state), the IFE can excite the AF-mode only. In
the discussion presented above, the only part proportional to
$b_{xyxy}L_xl_y$ gives an essential contribution to $\delta
\mathcal{W}_a(\bmL ,t)$. Using these relations, one can find that
all the terms do not affect the equation for $l_z$
(F-mode), whereas the equation for $l_y$ describing the
AF-mode acquires nonzero right-hand side and reads
as
\begin{equation}
\frac{d^2l_y}{dt^2}+ \omega^2_{\mathrm{AF}}l_y =  -2\gamma M_0
\frac{dH_z^{{\rm eff,} \sigma ^{\pm}}}{dt}+A(2\gamma M_0)^2h_y^{{\rm
lin} } \label{eq:sigma},
\end{equation}
where $h_y^{{\rm lin} }=- \partial \delta \mathcal{W}_a(\bmL
,t)/\partial l_y $ is the effective field. Then, after the
delta function substitution $\mathcal{E}(t)
\mathcal{E}^*(t) \to I_0\delta(t)$, we arrive at the following initial
conditions for this equation
\begin{gather}
\left(l_y\right)_{t=0} =  \pm \frac{\gamma M_0 f_{xyz}I_0^{\sigma
^{\pm}}}{2}\,, \\
\left(\frac{dl_y}{dt}\right)_{t=0} = -4\gamma ^2 AM_0^3 b_{xyxy}I_0^{{\rm lin}}\sin2\theta \label{eq:InCond},
\end{gather}
where $I_0^{\sigma ^{\pm}}$ and $I_0^{{\rm lin}}$ determine
independent action of circularly and linearly polarized light,
respectively, with $I_0^{\sigma^{\pm}}$ and $I_0^{{\rm lin}}$ being
the corresponding integrated pulse intensities. As one can see from
the equation, within the sigma-model approach the effective magnetic
field created by the IFE enters the equation through its time
derivative only, whereas the inertial mechanism is caused solely by
ICME.  Thus the \emph{field-derivative mechanism} of the action of
IFE, discussed previously for  compensated
antiferromagnets,~\cite{Galkin,Satoh1} is responsible for the
excitation of  spin oscillations in the $\Gamma_4$-phase of
dysprosium orthoferrite investigated here. We conclude that it is
difficult to realize the inertial mechanism  of the field pulse
action in the majority of orthoferrites at high temperatures where
the same $\Gamma_4$-phase is present. The inertial mechanism has
been observed  for a special phase of holmium orthoferrite where the
vector $\bmL$ is not collinear with the symmetry axis.~\cite{Kimel3}
On the other hand, for the present experiment the ICME leads to
inertial mechanism of the spin excitations.

After the action of the pulse,  only free spin oscillations
persist in the system. They are described by the solution
\begin{gather}
l_y^{\mathrm{free}}= a \cos(\omega_{\mathrm{AF}}t+\xi),  \\
m_z^{\mathrm{free}}= a \frac{\sqrt{d^2-Aq_1}}{A}\sin(\omega_{\mathrm{AF}}t+\xi) \label{eq:solution},
\end{gather}
where the amplitude $a$ and the phase $\xi$ are determined by the
initial conditions~\eqref{eq:InCond} as follows:
\begin{gather}
\tan \xi = \mp \frac{4AM_0b_{xyxy}I_0^{{\rm lin}}\sin2\theta
}{f_{xyz}I_0^{\sigma ^{\pm}}\sqrt{d^2-Aq_1}}, \\
a=\gamma M_0\sqrt{ \frac{ \left( f_{xyz}I_0^{\sigma
^{\pm}} \right)^2}{4}+\frac{\left( 2A b_{xyxy}M_0I_0^{{\rm
lin}}\sin2\theta\right)^2}{d^2-Aq_1}} \label{eq:a_xi}.
\end{gather}

Finally, we arrive at the previous result: if one of the two
mechanisms, IFE or ICME, is dominating, the the phase of the $m_z$
oscillations takes the values $\xi = 0$, $\pi$ or $\xi = \pm \pi/2$,
respectively. Thus, the observed time dependence of the Faraday
rotation oscillations is proportional to $ \sin
\omega_{\mathrm{AF}}t$ or  $ \cos \omega_{\mathrm{AF}}t$ for the
dominating role of IFE or ICME, respectively. If  none of the
mechanisms is truly dominating, then the observed phase should take
an intermediate value given by Eq.~(\ref{eq:a_xi}).

It is worth to note that the condition for domination of a certain
effect does not translate into a plain comparison of the effective
constant values $f_{xyz}$ and $2M_0b_{xyxy}$ for IFE and ICME,
respectively. The point is, the ICME contributes through the
inertial mechanism that is much more effective than the
field-derivative mechanism involved in the action of IFE. In our
calculation, this leads to appearance of the large multiplier
$A/\sqrt{d^2-Aq_1} =\gamma H_{\mathrm{ex}}/\omega_{\mathrm{AF}}$,
where $\gamma H_{\mathrm{ex}} \approx 20$~THz,
$H_{\mathrm{ex}}=2AM_0 \simeq $ 600~T is the exchange field of
orthoferrite,~\cite{Wijn} in the contribution of ICME, see
Eq.~\eqref{eq:a_xi}. Therefore the domination of IFE, for the same
value of the pulse fluence, needs at least 50 times higher value of
the corresponding constant, and the ratio $f_{xyz}/2M_0b_{xyxy}$ is
expected to be large enough for orthoferrites. Thus, the above
analysis gives us a possibility to estimate the values of constants
responsible for different inverse magneto-optical effects, IFE and
ICME.

\subsection{Comparison between the theory and the experiment}

In the previous discussion, based on the Landau-Lifshitz equations
and the nonlinear sigma model, we came to the conclusion that the
time dependence of $m_z$ induced via IFE and ICME is proportional to
$\sin \omega _{\rm AF}t$ and $\cos \omega _{\rm AF}t$, respectively.
The phase of the oscillation is constant and $m_z$ is proportional
to either $\sin \omega _{\rm AF}t$ or $\cos \omega _{\rm AF}t$ in
some region of the pump wavelength in Fig.~\ref{fig:fig7} (a). When
the pump pulse is in the visible region ($<$800~nm), the probe
polarization and $m_z$ oscillate as $\sin \omega _{\rm AF}t$. This
property is independent of temperature as shown in
Fig.~\ref{fig:fig8} (b). On the other hand, when the pump pulse is
in the near-infrared region (1000--1100~nm), the probe polarization
and $m_z$ oscillate as $\cos \omega _{\rm AF}t$. Thus, we can
conclude that the visible and near-infrared light pulses dominantly
induce spin precession via IFE and ICME, respectively.

A number of reasons can be given for why the dominant effect varies with pump wavelength.
IFE is induced by a pulse whose wavelength is near the transition $^6A_1 \rightarrow $ $^4T_2$ at 700 nm.
On the other hand, ICME is induced by a pulse whose wavelength is near the transition $^6A_1 \rightarrow $ $^4T_1$ at 1000 nm.
In addition, the Faraday rotation angle increases with decreasing wavelength in DyFeO$_3$.\cite{Tabor1,Chetkin}
This tendency agrees with the result for the IFE.

\section{CONCLUSIONS}

We have studied the dependence of photoinduced spin precession in
DyFeO$_3$ on the wavelength and polarization of a pump pulse with a
pump--probe magneto-optical technique.
The polarization rotation of the probe pulse was dependent on the
pump polarization.
Pulses propagating along the
$z$-axis with both circular and linear polarizations induced an
effective magnetic field (IFE and ICME) and spin precession.
The dominant component of the dielectric permittivity in both
effects was $\varepsilon _{xy}$, and IFE and ICME were induced by
its antisymmetric and symmetric parts $\varepsilon_{xy}^a$ and
$\varepsilon_{xy}^s$, respectively.

The phase and amplitude of the spin precession were dependent on the
pump wavelength in DyFeO$_3$.
A difference in the pump wavelength changes the dominant effect, giving rise to
the spin precession.  A visible pulse (wavelength $<$800~nm) induced the IFE, and the
oscillation of the probe polarization was proportional to $\sin \omega _{\rm
  AF}t$.  On the other hand, a near-infrared pulse (wavelength of 1000--1100~nm) induced the
ICME dominantly, and the oscillation was proportional to $\cos \omega _{\rm
  AF}t$.
When the pump wavelength was near the electron transition $^6A_1 \rightarrow $ $^4T_2$ at 700 nm and $^6A_1 \rightarrow $ $^4T_1$ at 1000 nm, the amplitude of the oscillation was larger than that of the other region.

The ratio of the effective magnetic fields via IFE and ICME,
$f_{xyz}/2M_0b_{xyxy}$, is expected to be large enough for
orthoferrites. However, the ellipticity of spin precession with
AF-mode is also so large. Therefore, even though linearly polarized
light pulse induces so weaker magnetic field than circularly
polarized one, ICME can give the same order contribution as IFE.

\section*{acknowledgments}
This work was supported by KAKENHI (19860020 and 20760008). B.~A.~I.
was partly supported by the grant No. 220-10 from the Ukrainian
Academy of Sciences and by the grant No.~5210 from STCU. We thank
A.~K.~Kolezhuk for useful discussions and help.

\newpage

\end{document}